\documentclass[aps,prb,twocolumn,floatfix,superscriptaddress,showpacs,a4paper,citeautoscript]{revtex4}

\usepackage{graphicx}
\usepackage{bbm,latexsym}
\usepackage{amsmath,amssymb}

\def\be{\begin{equation}}
\def\ee{\end{equation}}
\def\bea{\begin{eqnarray}}
\def\eea{\end{eqnarray}}

\usepackage{natbib}
\usepackage{color}

\newcommand{\etal}{\textit{et al.~}}
\newcommand{\abinitio}{\textit{ab initio~}}

\begin{document}
\title{Alloying effects on the optical properties of  Ge$_{1-x}$Si$_x$  nanocrystals from TDDFT and comparison with effective-medium theory}

\newcommand{\polytechnique}{Laboratoire des Solides Irradi\'es, \'Ecole Polytechnique, 
CNRS, and CEA-DSM, F-91128, Palaiseau, France}
\newcommand{\etsf}{European Theoretical Spectroscopy Facility (ETSF)}
\newcommand{\lyon}{Laboratoire de Physique de la Mati\`{e}re Condens\'{e} et Nanostructures,
Universit\'{e} Lyon I and CNRS, F-69622 Villeurbanne Cedex, France}

\author{Silvana Botti}
\affiliation{\polytechnique}
\affiliation{\lyon}
\affiliation{\etsf}

\author{Hans-Christian Weissker}
\affiliation{\polytechnique}
\affiliation{\etsf}

\author{Miguel A. L. Marques}
\affiliation{\lyon}
\affiliation{\etsf}

\date{\today}

\begin{abstract}
We present the optical spectra of Ge$_{1-x}$Si$_x$ alloy nanocrystals calculated with time-dependent density-functional theory in the adiabatic local-density approximation (TDLDA). The spectra change smoothly as a function of the composition $x$. On the Ge side of the composition range, the lowest excitations at the absorption edge are almost pure Kohn-Sham independent-particle HOMO-LUMO transitions, while for higher Si contents strong mixing of transitions is found. Within TDLDA the first peak is slightly higher in energy than in earlier independent-particle calculations. However, the absorption onset and in particular its composition dependence is similar to independent-particle results.  Moreover, classical  depolarization effects are responsible for a very strong suppression of the absorption intensity. We show that they can be taken into account in a simpler way using Maxwell-Garnett classical effective-medium theory. Emission spectra are investigated by calculating the absorption of excited nanocrystals at their relaxed geometry. The structural contribution to the Stokes shift is about 0.5\,eV. The decomposition of the emission spectra in terms of independent-particle transitions is similar to what is found for absorption. For the emission, very weak transitions are found in Ge-rich clusters well below the strong absorption onset.

\end{abstract}
\pacs{71.15.Mb, 78.67.-n, 78.67.Bf}
\maketitle

\section{Introduction}

The semiconductors silicon and germanium can form a substitutional
solid solution of the form Ge$_{1-x}$Si$_x$ covering the whole range
of compositions $x$. Pure Si is the most widely used material for
electronic applications since many years and its fabrication
technology is highly developed. However, the indirect band gap of bulk
Si presents a problem for light-emitting applications. A solution
that has been proposed to circumvent this problem is
nanostructurization of Si in structures comprising porous Si
\cite{ossicini-review}, nanowires \cite{nanowires-review}, as well as
Si nanocrystals. Moreover, ample use has been made of the fact that
Ge can  easily be combined with Si in
heterostructures. In addition, Ge nanocrystals in a matrix of SiO$_2$,
SiC, sapphire, or Si have been investigated by many
groups~\cite{takeoka-98-germanium,wilcoxon-01-germanium,schubert-02}. In
many of those cases, intermixing between Ge and Si is found in the
nanostructures.

The two materials show different properties upon
nanostructurization. While Si retains its character of an indirect
material \cite{weissker-free,weissker-structure,weissker-MSEB}, the work of several
groups showed that, depending on the structure, strain, etc., Ge
nanostructures can become quasi-direct, i.e., they exhibit very strong
transitions at or very close to the HOMO-LUMO transition
\cite{weissker-free,weissker-structure,weissker-MSEB,kholod-02,kholod-04,tsolakidis-05,melnikov-ssc}. The
different behavior reflects the different character of the band gaps
in the bulk materials. While both are indirect, the minimum gap in Si
lies between $\Gamma$ and a point near the X point, and the direct gap
is much larger. In Ge, the minimum gap between $\Gamma$ and L is
energetically very close to the direct gap at $\Gamma$. The effects of
confinement and structural relaxation result in a strong contribution
of the $\Gamma$--$\Gamma$ transition to the HOMO-LUMO transition, thus
resulting in short radiative lifetimes of Ge
clusters~\cite{weissker-MSEB,weissker-dscf}.

Therefore, when mixing both materials two questions arise: what is the
effect of the intermixing on the electronic properties of the
nanocrystals, and how are the different characters of the two materials
 combined. While the answer to these questions is essentially well
known for the bulk alloy, there are still few investigations
concerning the mixed nanostructures. Most experimental studies use
Stranski-Krastanov growth, (see, e.g., Ref.~~\onlinecite{schmidt-02}),
which results in relatively large structures in which it is safe to
assume that the effects of confinement and alloying act independently.
However, for smaller structures this cannot be taken for granted. For
small nanocrystals, photoluminescence experiments
\cite{takeoka-00-GeSi} have been compared with theoretical results
\cite{weissker-GeSi-prl}. Furthermore, Ge-Si nanowires
have been investigated experimentally \cite{yang-06} and theoretically
\cite{migas-07}.

Previous theoretical studies have focused on the interplay of
confinement and alloying \cite{weissker-GeSi-prl,migas-07}. However,
these \abinitio\ calculations were performed within the
independent-particle approximation based on the Kohn-Sham scheme of
\textit{static} density-functional theory (DFT). Therefore, important
many-body effects have been neglected, viz., the self-energy
corrections describing the effect of the excitation of the electrons
or holes individually, as well as the electron-hole
interaction. (Nonetheless, these two effects are found to cancel each
other to a large extent for many systems.\cite{delerue-00}\phantom{})
Furthermore, these calculations miss the very important depolarization
or crystal local-field effects.

The way we choose to improve upon the independent-particle results is
provided by time-dependent DFT~\cite{runge-gross-84,pMarques2006,botti-review-07} (TDDFT) in
which the many-body effects neglected in static DFT are introduced by
the so-called exchange-correlation kernel $f_{\textrm{xc}}$. With respect
to the independent-particle results, the excitation energies are
corrected, and the transitions between the independent-particle states
are mixed. The degree of the mixing indicates the degree to which the
independent-particle approximation fails to provide a good description
of the system.

Within TDDFT, we use the adiabatic local-density approximation (ALDA,
also known as TDLDA) of the exchange-correlation kernel. The choice is
motivated by the fact that TDLDA yields good results for optical
spectra of isolated systems~\cite{pMarques2001} as well as for non-zero
momentum transfers~\cite{weissker-ixs-prl}. Note, however, that the
ALDA is well known to fail in some cases, the most important of which are perhaps
extended systems~\cite{rmp_lucia,botti-review-07}. Within TDDFT, the
random-phase approximation (RPA) is obtained by neglecting the
exchange-correlation kernel, i.e., by setting $f_{\text{xc}}=0$. From this, the
independent-particle approximation results from the neglect of the microscopic terms of
the variation of the Hartree potential. In other words, the difference between the independent-particle
approximation and the RPA are the depolarization effects which are due to the
inhomogeneity of the system.

The simplest way to account for depolarization effects within an approximated classical picture is
to apply the Maxwell-Garnett effective-medium theory to the complex dielectric function of the bulk alloy
crystals. We applied the effective medium theory to our alloy clusters in order to check if the depolarization effects are dominant with respect to confinement effects and further many-body 
corrections. In fact, if it is the case,  the spectra obtained using the effective-medium theory are in good agreement with full TDLDA results. These model calculations thus give a good overall description of the
optical spectra at a much lower computational cost than TDLDA.

In the present paper we first present TDLDA results for Ge$_{1-x}$Si$_x$
nanocrsytals, focusing on alloying effects. 

The results are then compared with the previous independent-particle calculations of Ref.~~\onlinecite{weissker-GeSi-prl}, highlighting the effects of the depolarization, as well as of the mixing of transitions. We will also consider depolarization effects alone, decoupled from confinement effects, by applying  Maxwell-Garnett classical effective-medium theory. 
Finally, the emission properties of the nanocrystals are investigated by considering the geometries obtained after excitation of an electron-hole pair.


\begin{figure}[t]
  \centering
  \includegraphics[width=.5\columnwidth,angle=-0]{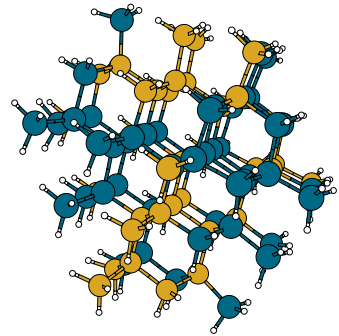}
  \caption{ \label{fig:model} (Color online) Example of the studied nanocrystal
  structures: Ge$_{48}$Si$_{35}$H$_{108}$. The colors are: Si (yellow),
  Ge (blue), and H (white).}
\end{figure}


\begin{figure*}
  \centering
  \includegraphics[width=.95\columnwidth,angle=-0]{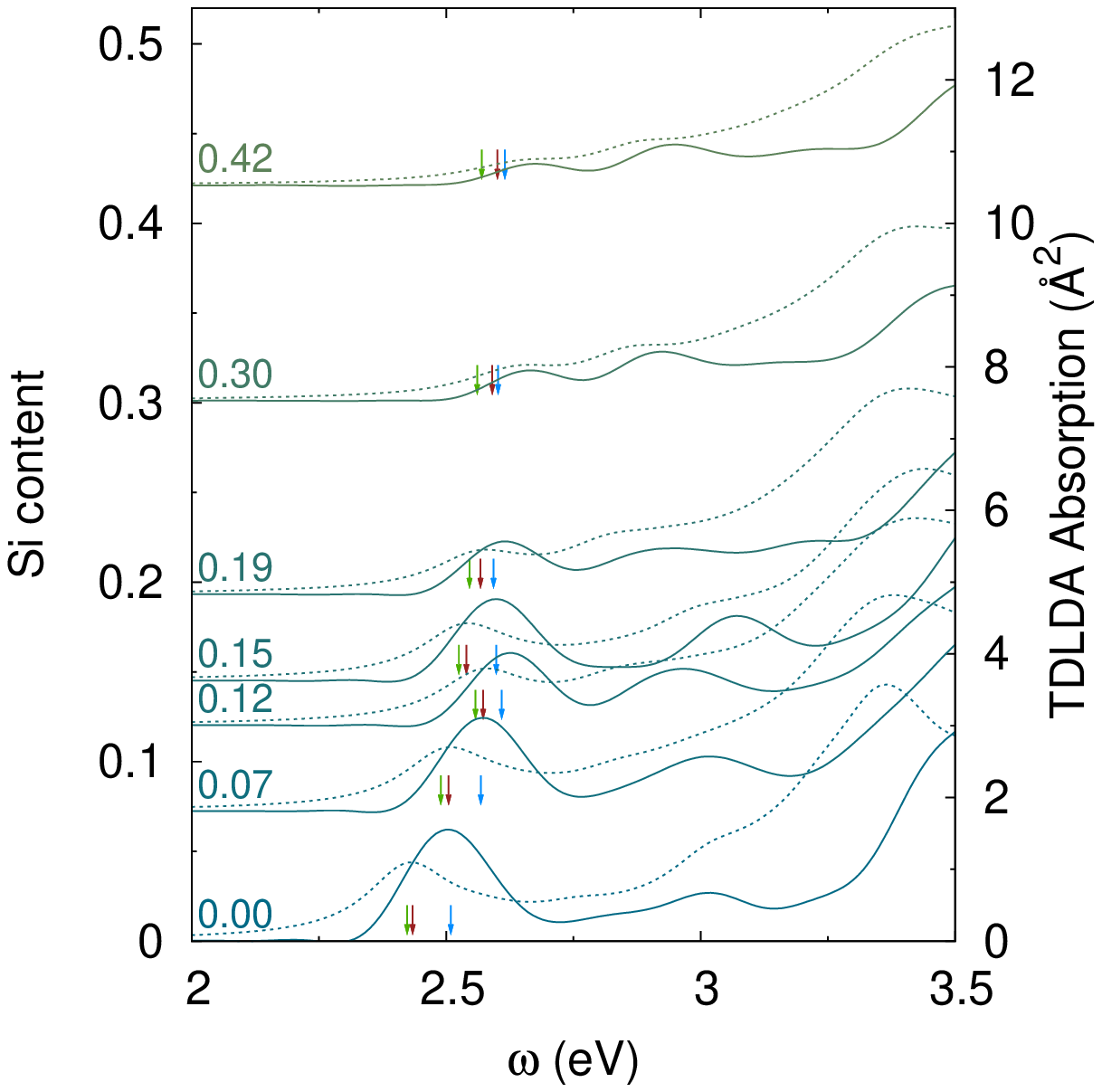}
  \includegraphics[width=.95\columnwidth,angle=-0]{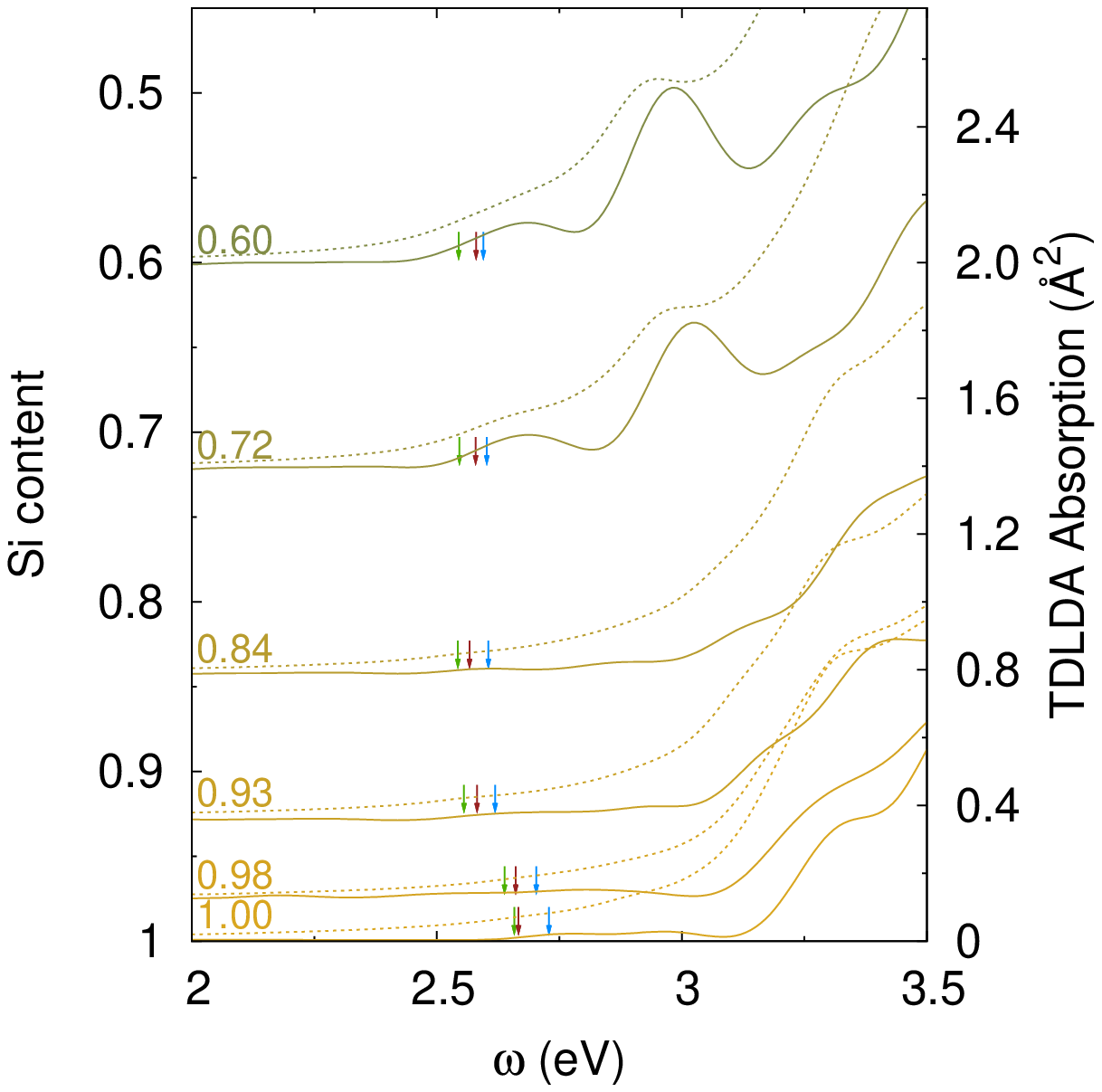}
  \caption{\label{fig:abs_IP} (Color online) Absorption spectra as a
  function of the Si content of the clusters:
  Independent-particle spectra (dashed lines) compared with TDLDA
  results (solid lines). The green arrows (arrows at lowest energy) mark the HOMO-LUMO gap,
  while the red arrows (middle arrows) mark the $\Delta$SCF excitation energies and the blue arrows 
  (arrows at highest energy) mark the first transitions of the Casida's analysis. The
  independent-particle curves are divided by a factor of 15. Note the different scale in the two panels.}
\end{figure*}

\section{Calculation details}
\label{section:calculations}


We used the same modeling scheme as used in
Ref.~~\onlinecite{weissker-GeSi-prl} considering Ge$_{1-x}$Si$_x$ nanocrystals with a fixed size made of 83 Ge and Si atoms. Quasi-spherical nanocrystals were
built starting from one atom and adding nearest neighbors shell
by shell, assuming bulk-like tetrahedral coordination. The outer
bonds were saturated by hydrogen atoms. Alloying between Ge and Si was
introduced by randomly exchanging Ge atoms by Si. The surface was then
passivated with H atoms to saturate the remaining dangling
bonds.

In Ref.~~\onlinecite{weissker-GeSi-prl}, the study of Ge$_{1-x}$Si$_x$ nanocrystals with
the same number of atoms for ten different atomic configurations
demonstrated that those with nearly uniformly distributed Ge and Si atoms possess the
lowest total energies and nearly equal excitation energies. On the
other hand, nanocrystals with deliberately clustered Si atoms and,
hence, rather different excitation energies give rise to total
energies substantially higher than the average. As their probability of
occurrence is, consequently, small, the configurational average can be replaced by
the study of only one nanocrystal with nearly uniformly distributed Si
and Ge atoms for each composition $x$.  We selected nanocrystals with a
total number of 83 atoms of Ge and Si, having a diameter of about
1.5\,nm. This radius corresponds to a sphere of the volume occupied by 83 atoms in the bulk. These nanocrystals are large enough to exhibit the characteristics of
nanocrystals as opposed to much smaller structures which show a
molecular-like behavior
\cite{weissker-free,weissker-emt,weissker-dscf}. Moreover, they are 
small enough to present significant confinement effects on the electronic
states. An example of the atomic arrangement 
is shown in Fig.~\ref{fig:model} for Ge$_{48}$Si$_{35}$H$_{108}$.

To obtain the relaxed geometries of the  Ge$_{1-x}$Si$_x$  nanocrystals, we used the
plane-wave code VASP \cite{web-vasp} within the local-density
approximation (LDA) in the parametrization of Perdew and Zunger
\cite{perdew-zunger} and the projector-augmented wave (PAW) method
\cite{bloechl-94}. This computational set-up is the same as the one
employed in Ref.~\onlinecite{weissker-GeSi-prl}.

Starting from the relaxed geometries we obtained the optical spectra at zero
temperature using TDDFT as implemented in the computer code {\tt
octopus}~\cite{octopus,octopus2}. The electron-ion interaction is
described through norm-conserving pseudopotentials~\cite{troullier91}
and the LDA~\cite{perdew-zunger} is
employed in the adiabatic approximation for the xc potential (i.e., we
applied the TDLDA). The time-dependent Kohn-Sham equations, 
in this code, are represented in a real-space regular grid,
using a spacing of 0.275\,\AA\, at which the calculations are converged. The simulation box is
constructed by joining spheres of radius 4.5\,\AA, centered around
each atom.

To calculate the optical response we excite the system from its ground
state by applying a delta electric field $E_0
\delta(t)$. The real-time response to this perturbation is Fourier
transformed to get the dynamical polarizability $\alpha(\omega)$ in
the frequency range of interest. The absorption cross section
$\sigma(\omega)$, is then obtained from the relation:
\begin{equation}
  \sigma (\omega) = \frac{4 \pi}{c} \omega\; \mathfrak{Im}\: {\alpha(\omega)}
  \,,
\end{equation}
where $c$ is the velocity of light in vacuum.
For a technical description of this method we refer to Refs.~~\onlinecite{octopus} and~~\onlinecite{castro04}. To obtain the RPA spectra in this approach we just kept the
exchange-correlation potential  fixed during the propagation (which amounts
to making $f_{\rm xc}=0$). The independent-particle spectra were calculated by
further fixing the Hartree potential to its initial value. A time step
of 0.0055\,$\hbar$/eV and a total propagation time of 37.5\,$\hbar$/eV
were sufficient to ensure a stable propagation. We estimate our
numerical precision in the spectra to be better than 0.05\,eV.  The
mixing of independent-particle transitions in the spectra has
subsequently been investigated using the Casida's formulation of
linear-response TDDFT~\cite{casida-95,casida-96}.

In order to test the coherence of our calculations, we compared
independent-particle spectra computed using matrix elements calculated with the VASP code
\cite{adolph-01,weissker-tetra,weissker-free} to spectra obtained by time
propagation by means of the {\tt octopus} code. No substantial differences were
found between the results.

Due to the interest in the luminescence of this kind of systems, the lowest
excitation energies are of particular importance. As a consequence, we
decided to compare three
different quantities. Besides the lowest excitations of the TDLDA absorption
spectra, we present the Kohn-Sham HOMO-LUMO gap of the ground state
and the $\Delta$SCF excitation energies calculated as the difference
$E_{\Delta\textrm{SCF}} = E(N,\textrm{e--h}) - E(N)$, where $E(N)$ is the total energy of the ground-state
and $E(N,\textrm{e--h})$ is the total energy of the excited-state
configuration where one electron has been promoted from the HOMO into
the LUMO. In this way, an excited configuration is modeled and the electron-hole interaction is partially taken into account. This approach enables also a subsequent ionic relaxation with the electron-hole pair present, which yields the description of the excited-state geometries. Using the excited-state geometries
we were also able to evaluate emission spectra.


\section{Results}
\label{section:results}

\subsection{Ground-state geometries: absorption}

Results for the photoabsorption of the Ge$_{1-x}$Si$_x$ nanocrystallites are presented in Fig.~\ref{fig:abs_IP}.
In the figure we compare the independent-particle response (dashed lines) with full TDLDA calculations
(solid lines) for the whole range of compositions. The dependence of the curves on the composition turns out to be quite smooth.
When going from Ge-rich clusters to Si-rich clusters, we observe a shift to higher energies of the onset and a suppression of the absorption strength of the first peak (note the different scales of both panels). Moreover, the dependence of the peak position on the composition $x$ is roughly the same in the independent-particle and the TDLDA schemes. For the clusters studied, the total intensity of the independent-particle spectra is strongly suppressed in TDLDA (the independent particle curves are divided by a factor of 15 in Fig.~\ref{fig:abs_IP}).
Indeed, this quenching of the absorption is a well known effect 
that is due to the inclusion of classical depolarization effects, and not from the
exchange-correlation effects accounted for by the TDLDA kernel. This can be verified by calculating the absorption spectrum within the RPA, which includes all classical effects due to the variations of the Hartree potential but neglects quasi-particle and excitonic effects. Indeed, the spectra we calculated within RPA are so close to the TDLDA spectra shown in Fig.~\ref{fig:abs_IP} that we chose not to show them.

\subsection{Transition analysis and excitation energies}

For Ge-rich nanoparticles, both the position and the composition dependence of the peaks are already well described at the level of the independent-particle approximation. This perhaps surprising fact can be explained in terms
of compensation of quasiparticle corrections
and binding energies of the excitons.
For small $x$ at the absorption edge, the first peak
is strong and appears essentially  at the HOMO-LUMO energy gap.
The absorption edge has a completely different nature in Si-rich clusters. In fact, in this case the 
peaks of lowest energy have a vanishing oscillator strength in TDLDA,
thereby blue-shifting the absorption edge with respect to the HOMO-LUMO gap.

In order to analyze the origin of the different peaks in the spectra as a function of the composition of the  Ge$_{1-x}$Si$_x$  alloy,
we decomposed the excitations in sums of Kohn-Sham particle-hole transitions through the solution of  Casida's equation
\cite{casida-95,casida-96}.
We found that on the Ge side of the composition range, the lowest peak of the spectra which defines the absorption edge
is produced essentially by a strong, pure transition between the Kohn-Sham HOMO and LUMO. This is certainly one more reason for the similarity between the independent-particle and the TDLDA spectra.
The large peak with HOMO-LUMO character decreases in intensity with increasing percentage of Si in the  Ge$_{1-x}$Si$_x$  alloy, until the composition of about $x=0.2$ when it disappears. For even smaller $x$ the absorption at the onset is determined by a strong mixture of transitions between states close to HOMO and LUMO. However, the very lowest transitions are forbidden, producing a significant blue shift of the absorption edge. In the intermediate energy range and for all compositions, excitations can be decomposed as a sum of many contributions. 

In Fig.~\ref{fig:abs_IP} we also show the HOMO-LUMO gap (green arrows), the lowest excitation calculated within the $\Delta$SCF approximation (red arrows), and the first excitation within TDLDA (blue arrows, these excitations are dark on the Si side of the composition). For all cases we find that the first $\Delta$SCF excitation is, as expected, blue-shifted with respect to the HOMO-LUMO gap, and that the first TDLDA transition is at slightly higher energies. The differences are, however, quite small, and of the order of one tenth of an eV. This is known for this class of systems~\cite{degoli-04}, part of it being due to the cancellation of self-energy and excitonic effects \cite{delerue-00}. However, the differences appear to be slightly larger in the region of intermediate composition $x$, i.e., of a greater degree of structural disorder. Similarly, Degoli \etal found in Si nanocrystals that the differences become larger in cases of stronger 
localization~\cite{degoli-04}.
It is also clear from the plot how the first transition becomes forbidden while going from Ge-rich to Si-rich clusters --- this behavior is reminiscent of the different character of the band gap in the parent bulk Ge and Si.

\subsection{Absorption from classical effective-medium theory}
\label{section:EMT}

\begin{figure}
  \centering
  \includegraphics[width=.95\columnwidth,angle=-0]{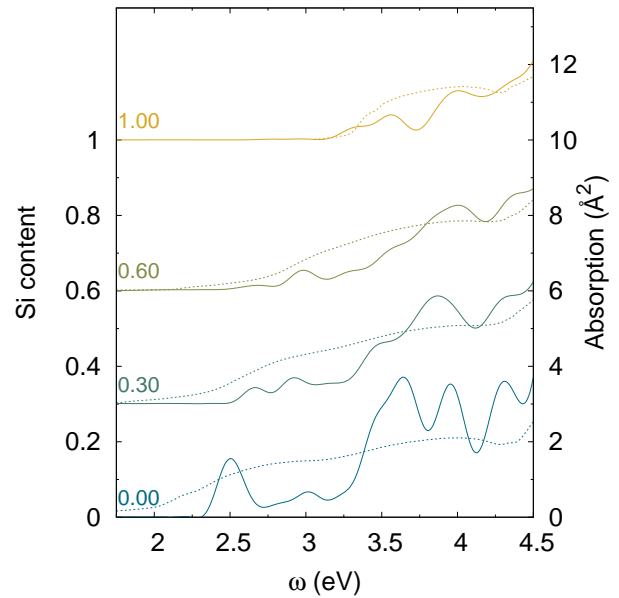}
  \caption{\label{fig:abs_EMT} (Color online) Absorption spectra as a
  function of the Si content of the clusters:
  Effective-medium theory (dashed lines) using the dielectric function of bulk  Ge$_{1-x}$Si$_x$ , compared with TDLDA
  results (solid lines).}
\end{figure}

A much simpler approach to model the absorption cross section of a Ge$_{1-x}$Si$_x$
nanocrystal is to start from the complex dielectric function of the
corresponding bulk alloy crystal and to apply the effective-medium theory~\cite{emt,emt-b}.
This classical approach is based on Maxwell's
equations and neglects completely the microscopic details, such as atoms and bonds.
Of course, this assumption is better justified when the size of the
system is large. However, it always handles correctly the boundary conditions 
for the Maxwell's equations at the
interfaces, which give the very important contributions to the dielectric
response through the classical depolarization effects. Often, these classical contributions are enough to describe the physics of the dielectric response of a  composite system made of objects embedded in some matrix~\cite{sottile05,botti02,botti04}.  
Our clusters can be
considered as a family of spheres of volume $V_{\textrm{obj}}$ cut from a Ge$_{1-x}$Si$_x$ bulk alloy.
The Maxwell-Garnett expression~\cite{emt} 
in the specific case of an isolated spherical object in vacuum yields:
\renewcommand{\Im}{{\mathfrak{Im}\:}}
\renewcommand{\Re}{{\mathfrak{Re}\:}}
\begin{equation}
 \sigma (\omega)  = 9 \frac{\omega}{c} V_{\textrm{obj}}
\frac{\Im \epsilon(\omega)}{\left[ \Re \epsilon( \omega) +2 \right]^2 +
\left[ \Im \epsilon(\omega) \right]^2}  \,,
\end{equation}
where $\epsilon$ is the experimental complex dielectric
function of the bulk alloy. 
To represent well the extension of the polarizable nanocrystals, we took an average distance of the furthermost saturating hydrogen atoms  to obtain the radius of the cluster, and consequently $V_{\textrm{obj}}$. This results in a radius slightly larger than the one mentioned above which takes into account only the Ge and Si atoms. We used the value of $R_\textrm{obj}=9.1$\,\AA, but the results are fairly insensitive to a (reasonable) choice of this value.
The experimental dielectric function of the bulk alloy with precisely the needed composition $x$ has been obtained from a discrete set of measurements\cite{humlicek-89} using a recently developed interpolation scheme\cite{weissker-interpolation}. This scheme interpolates $\Im \epsilon(\omega)$ making use of the screening sum rule and the positive definiteness of the spectra. $\Re \epsilon(\omega)$ was subsequently obtained by means of the Kramers-Kronig relations after fitting appropriate tails to the imaginary parts. The latter procedure was tested on the input curves to insure the quality of the real part.

In Fig.~\ref{fig:abs_EMT} we show spectra calculated using the effective-medium theory for selected compositions. Comparing to the first-principles TDLDA curves, we can see that, already for this relatively small size of clusters, the classical theory gives a quite good overall description of the absorption spectrum. As expected, effective-medium theory is not capable of describing the peaks of the individual transitions, but it describes correctly the intensity and the trends of the spectrum. Once again these results confirm that the dependence on the composition of the optical spectra is smooth and that the confinement and alloying effects act independent to a large degree. In fact, within this classical scheme, only the alloying effects determine the variation of the absorption response as a function of $x$. 
In particular, the confinement-induced opening of the HOMO-LUMO gaps and the resulting blue-shift of the absorption onset are not accounted for. The confinement effects could be described by introducing a size-dependent crystallite dielectric function which can then be used to calculate the spectra of the crystallites in a different environment.\cite{weissker-emt} In this sense, comparison of the TDLDA results with the present effective-medium approach gives an idea of the importance of the confinement effects on the overall spectra.

We note that these model calculations can be performed at negligible computational cost, and therefore provide a simple and fast method to obtain reasonable spectra for medium and large nanocrystallites.

Given the strong influence of the depolarization effects, the question arises as to why the independent-particle spectra have been successfully compared previously with experiment~\cite{weissker-free}. This agreement is, in fact, not a fortuitous coincidence, but due to the experimental conditions. The experiment has been done on Ge NCs inside a matrix of sapphire\cite{tognini-96}. This reduces strongly the depolarization effects because it reduces the inhomogeneity of the system. The calculations, on the other hand, treated NCs in vacuum but neglected the depolarization effects. Therefore, together with the cancellation between self-energy effects and electron-hole interaction mentioned above, the independent-particle approximation provides in fact a good description of the spectra of this particular experiment.


\subsection{Excited-state geometries: emission}
\label{section:excited}


\begin{figure*}
  \centering
  \includegraphics[width=.95\columnwidth,angle=-0]{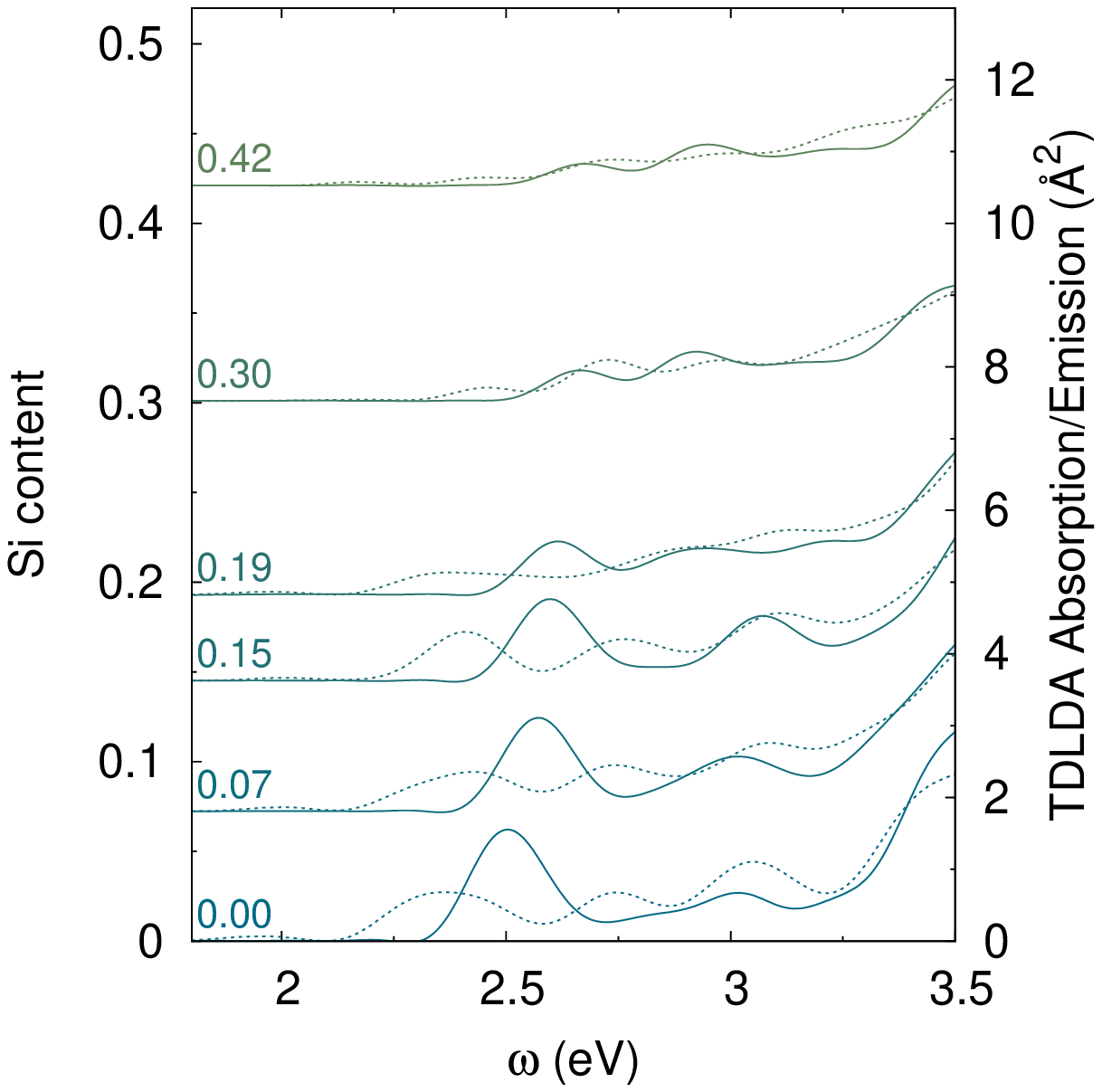}
  \includegraphics[width=.95\columnwidth,angle=-0]{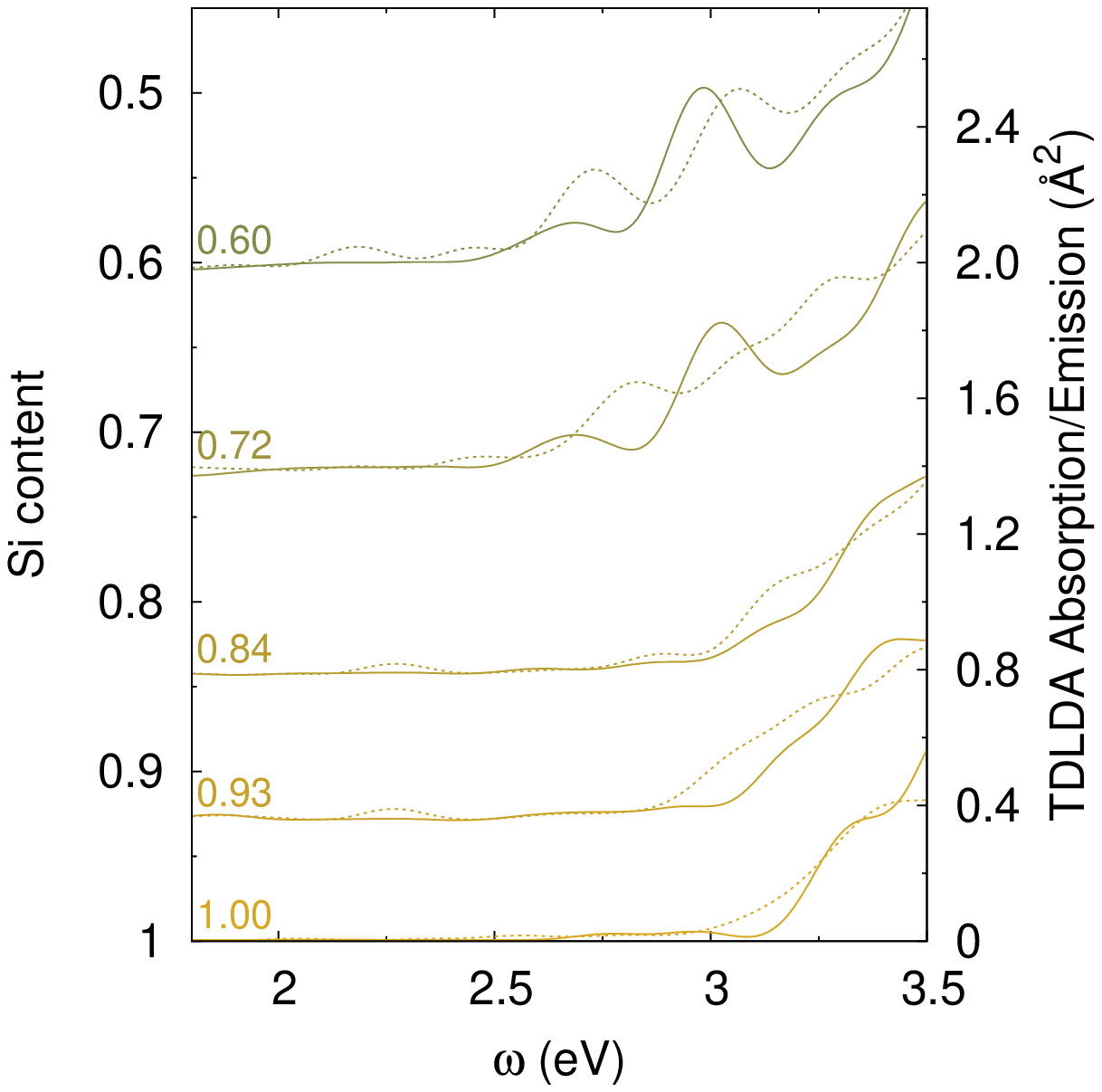}
  \caption{\label{fig:abs_em} (Color online) Absorption (solid
  lines) and emission (dashed lines) spectra as a function of the
  Si content of the clusters. Note the different scales.
  }
\end{figure*}


In order to calculate the emission properties, we consider the geometry of the relaxed nanocrystals where the ionic relaxation has been carried out after transferring an electron from the HOMO to the LUMO Kohn-Sham orbital. As the radiative lifetimes are usually much longer than the times that the electrons (holes) take to relax to the LUMO (HOMO), we can assume thermalized electron-hole pairs. As their lifetimes are then determined by the exponential factor describing their distribution~\cite{dexter-58,delerue-93}, it is the onset of the emission which reflects the emission properties, while the higher parts of the spectrum are suppressed.

Stimulated emission spectra can be easily obtained within our formalism by calculating the absorption cross-section  $\tilde \sigma_\textrm{abs}(\omega)$ at the excited-state geometry. Luminescence spectra can then be calculated from the van Roosbroeck-Shockley model~\cite{vRS}
\begin{equation}
\label{eq:luminescence}
  \sigma_\textrm{lum}(\omega) \sim \omega^2 \frac{1}{e^{\hbar \omega/k_\textrm{b}T}-1} \tilde\sigma_\textrm{abs}(\omega)
  \,,
\end{equation}
where $k_\textrm{b}$ is the Boltzmann constant and $T$ is the temperature. 

In Fig.~\ref{fig:abs_em} we compare the absorption cross sections for the ground-state (solid lines) and the excited-state (dashed lines) geometries for  Ge$_{1-x}$Si$_x$  clusters in all the composition range. The energy difference between the onsets of absorption and emission corresponds to the structural contribution to the Stokes shift.

\begin{figure*}
  \centering
  \includegraphics[width=.95\columnwidth,angle=-0]{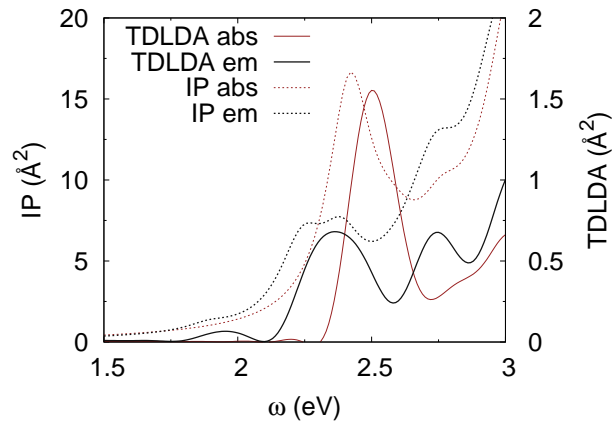}
   \caption{\label{fig:abs_em_IP_TDLDA} (Color online) Absorption (red) and emission (black) spectra of the pure Ge nanocrystal in the independent-particle approximation (dashed) and the TDLDA (solid). Note the different scales.
  }
\end{figure*}

In Fig. \ref{fig:abs_em_IP_TDLDA} we compare the independent-particle result with the TDLDA for the absorption  $\tilde \sigma_\textrm{abs}(\omega)$ at the excited-state geometry. The conclusions with respect to their similarity  drawn  for the ground state remain valid for the excited-state geometries. The same applies to the transition analysis using Casida's equation. The character of the emission onset on the Ge side of the composition range is already well described within the independent-particle approximation. The decomposition of the excitations as a sum of Kohn-Sham transitions provides a picture strictly analogous to the one for the absorption spectra: the lowest transitions of the Ge-rich nanocrystals correspond to almost pure Kohn-Sham transitions, while for large $x$ and for the higher transitions, independently of $x$, strong mixing is found.

However, it is important to note that a very weak peak appears at about 1.95\,eV. This is red-shifted by about 0.5~eV with respect to the ground-state calculation. The peak, which can be easily seen in Fig.~\ref{fig:abs_em_IP_TDLDA}, is clearly present in both the independent-particle result and the TDLDA result. It occurs for all compositions on the Ge side of the $x$ range and is almost composition independent. It appears to be connected with the lowering of the symmetry as compared to the ground state where the pure Si or Ge nanocrystals without alloying have T$_d$ symmetry and the HOMO-LUMO transition is threefold degenerate. The symmetry breaking due to the geometry relaxation under excitation has therefore a much stronger effect than the introduction of the alloying, which at the Ge-rich side splits the degeneracy only slightly and which does not change the character of the strong absorption onset at the HOMO-LUMO transition.
Due to the argument above, cf. Eq.~\ref{eq:luminescence}, the appearance of this weak peak will strongly increase the radiative lifetimes of the systems. We conjecture that this might be responsible for the fact that even though several theoretical predictions coincide in that Ge nanostructures should have strong transitions at the absorption onset, few experiments have been able to detect luminescence from excitons in Ge nanocrystals.


\section{Conclusions}

The absorption spectra of free Ge$_{1-x}$Si$_x$ have been calculated within time-dependent density-functional theory in the adiabatic local-density approximation. The changes of the spectra upon changing composition $x$ are smooth. In particular at the absorption onset, the position and the composition dependence of the spectra is found to be already well represented by independent-particle results. The analysis of the solutions of Casida's equation show that this is due to the fact that the first transition of Ge-rich nanocrystals corresponds to an almost pure independent-particle transition. The TDLDA onsets are slightly blue-shifted with respect to their independent-particle counterpart, their composition dependence at the Ge side of the compositional range is practically the same. For higher Si contents, a mixing of many independent-particle transitions is found.

Depolarization effects are strong and their inclusion alone, even within a simplified classical model, on top of an independent-particle calculation allows to get the correct physical picture of the optical response. They can be approximately taken into account at a negligible computational cost using Maxwell-Garnett effective-medium theory. 

Further many-body terms do not modify significantly the spectra due to the cancellation of opposite contributions given by quasiparticle corrections and excitonic effects.
As a consequence, the effects of the excitation of an electron-hole pair do not alter the comparison between the independent-particle spectra and the TDLDA results. 

Emission spectra have been investigated using the geometry of excited nanocrystals. A Stokes shift of about 0.5\,eV is found. Very weak peaks appear at the absorption onset for all systems on the Ge side which suppress the radiative transition probability and lead to long radiative lifetimes.



\section{Acknowledgments}

The authors would like to thank Lucia Reining for many fruitful
discussions.  All TDDFT calculations were performed at the
Laborat\'{o}rio de Computa\c{c}{\~a}o Avan\c{c}ada of the University
of Coimbra. The authors were partially supported by the EC Network of
Excellence NANOQUANTA (NMP4-CT-2004-500198) and the ETSF e-I3
(INFRA-2007-211956). S.\ Botti acknowledges financial support from
French ANR (JC05\_46741 and NT05-3\_43900).  H.-Ch.~W. acknowledges support from the European Union through the individual Marie Curie Intra-European Grant No. MEIF-CT-2005-025067.
M. A. L. Marques acknowledges partial support from the Portuguese FCT through the project
PTDC/FIS/73578/2006 and from the French ANR (ANR-08-CEXC8-008-01).



\begin{thebibliography}{49}
\expandafter\ifx\csname natexlab\endcsname\relax\def\natexlab#1{#1}\fi
\expandafter\ifx\csname bibnamefont\endcsname\relax
  \def\bibnamefont#1{#1}\fi
\expandafter\ifx\csname bibfnamefont\endcsname\relax
  \def\bibfnamefont#1{#1}\fi
\expandafter\ifx\csname citenamefont\endcsname\relax
  \def\citenamefont#1{#1}\fi
\expandafter\ifx\csname url\endcsname\relax
  \def\url#1{\texttt{#1}}\fi
\expandafter\ifx\csname urlprefix\endcsname\relax\def\urlprefix{URL }\fi
\providecommand{\bibinfo}[2]{#2}
\providecommand{\eprint}[2][]{\url{#2}}

\bibitem[{\citenamefont{Bisi et~al.}(2000)\citenamefont{Bisi, Ossicini, and
  Pavesi}}]{ossicini-review}
\bibinfo{author}{\bibfnamefont{O.}~\bibnamefont{Bisi}},
  \bibinfo{author}{\bibfnamefont{S.}~\bibnamefont{Ossicini}}, \bibnamefont{and}
  \bibinfo{author}{\bibfnamefont{L.}~\bibnamefont{Pavesi}},
  \bibinfo{journal}{Surf. Sci. Rep.} \textbf{\bibinfo{volume}{38}},
  \bibinfo{pages}{1} (\bibinfo{year}{2000}).

\bibitem[{\citenamefont{Lu and Lieber}(2006)}]{nanowires-review}
\bibinfo{author}{\bibfnamefont{W.}~\bibnamefont{Lu}} \bibnamefont{and}
  \bibinfo{author}{\bibfnamefont{C.~M.} \bibnamefont{Lieber}},
  \bibinfo{journal}{J. Physics D: Appl. Phys.} \textbf{\bibinfo{volume}{39}},
  \bibinfo{pages}{R387} (\bibinfo{year}{2006}).

\bibitem[{\citenamefont{Takeoka et~al.}(1998)\citenamefont{Takeoka, Fujii,
  Hayashi, and Yamamoto}}]{takeoka-98-germanium}
\bibinfo{author}{\bibfnamefont{S.}~\bibnamefont{Takeoka}},
  \bibinfo{author}{\bibfnamefont{M.}~\bibnamefont{Fujii}},
  \bibinfo{author}{\bibfnamefont{S.}~\bibnamefont{Hayashi}}, \bibnamefont{and}
  \bibinfo{author}{\bibfnamefont{K.}~\bibnamefont{Yamamoto}},
  \bibinfo{journal}{Phys. Rev. B} \textbf{\bibinfo{volume}{58}},
  \bibinfo{pages}{7921} (\bibinfo{year}{1998}).

\bibitem[{\citenamefont{Wilcoxon et~al.}(2001)\citenamefont{Wilcoxon,
  Provencio, and Samara}}]{wilcoxon-01-germanium}
\bibinfo{author}{\bibfnamefont{J.~P.} \bibnamefont{Wilcoxon}},
  \bibinfo{author}{\bibfnamefont{P.~P.} \bibnamefont{Provencio}},
  \bibnamefont{and} \bibinfo{author}{\bibfnamefont{G.~A.}
  \bibnamefont{Samara}}, \bibinfo{journal}{Phys. Rev. B}
  \textbf{\bibinfo{volume}{64}}, \bibinfo{pages}{035417}
  (\bibinfo{year}{2001}).

\bibitem[{\citenamefont{Schubert et~al.}(2002)\citenamefont{Schubert, Kaiser,
  Hedler, Wesch, Gorelik, Glatzel, lich, Wunderlich, He\ss, and
  Goetz}}]{schubert-02}
\bibinfo{author}{\bibfnamefont{C.}~\bibnamefont{Schubert}},
  \bibinfo{author}{\bibfnamefont{U.}~\bibnamefont{Kaiser}},
  \bibinfo{author}{\bibfnamefont{A.}~\bibnamefont{Hedler}},
  \bibinfo{author}{\bibfnamefont{W.}~\bibnamefont{Wesch}},
  \bibinfo{author}{\bibfnamefont{T.}~\bibnamefont{Gorelik}},
  \bibinfo{author}{\bibfnamefont{U.}~\bibnamefont{Glatzel}},
  \bibinfo{author}{\bibfnamefont{J.~K.} \bibnamefont{lich}},
  \bibinfo{author}{\bibfnamefont{B.}~\bibnamefont{Wunderlich}},
  \bibinfo{author}{\bibfnamefont{G.}~\bibnamefont{He\ss}}, \bibnamefont{and}
  \bibinfo{author}{\bibfnamefont{K.}~\bibnamefont{Goetz}}, \bibinfo{journal}{J.
  Appl. Phys.} \textbf{\bibinfo{volume}{91}}, \bibinfo{pages}{1520}
  (\bibinfo{year}{2002}).

\bibitem[{\citenamefont{{H.-Ch. Weissker} et~al.}(2002)\citenamefont{{H.-Ch.
  Weissker}, Furthm\"uller, and Bechstedt}}]{weissker-free}
\bibinfo{author}{\bibnamefont{{H.-Ch. Weissker}}},
  \bibinfo{author}{\bibfnamefont{J.}~\bibnamefont{Furthm\"uller}},
  \bibnamefont{and}
  \bibinfo{author}{\bibfnamefont{F.}~\bibnamefont{Bechstedt}},
  \bibinfo{journal}{Phys. Rev. B} \textbf{\bibinfo{volume}{65}},
  \bibinfo{eid}{155328} (\bibinfo{year}{2002}).

\bibitem[{\citenamefont{{H.-Ch. Weissker}
  et~al.}(2003{\natexlab{a}})\citenamefont{{H.-Ch. Weissker}, Furthm\"uller,
  and Bechstedt}}]{weissker-structure}
\bibinfo{author}{\bibnamefont{{H.-Ch. Weissker}}},
  \bibinfo{author}{\bibfnamefont{J.}~\bibnamefont{Furthm\"uller}},
  \bibnamefont{and}
  \bibinfo{author}{\bibfnamefont{F.}~\bibnamefont{Bechstedt}},
  \bibinfo{journal}{Phys. Rev. B} \textbf{\bibinfo{volume}{67}},
  \bibinfo{eid}{245304} (\bibinfo{year}{2003}{\natexlab{a}}).

\bibitem[{\citenamefont{{H.-Ch. Weissker}
  et~al.}(2003{\natexlab{b}})\citenamefont{{H.-Ch. Weissker}, Furthm\"uller,
  and Bechstedt}}]{weissker-MSEB}
\bibinfo{author}{\bibnamefont{{H.-Ch. Weissker}}},
  \bibinfo{author}{\bibfnamefont{J.}~\bibnamefont{Furthm\"uller}},
  \bibnamefont{and}
  \bibinfo{author}{\bibfnamefont{F.}~\bibnamefont{Bechstedt}},
  \bibinfo{journal}{Mat. Sci. Eng. B} \textbf{\bibinfo{volume}{101}},
  \bibinfo{pages}{39} (\bibinfo{year}{2003}{\natexlab{b}}).

\bibitem[{\citenamefont{Kholod et~al.}(2002)\citenamefont{Kholod, Ossicini,
  Borisenko, and Arnaud~d\char39{}Avitaya}}]{kholod-02}
\bibinfo{author}{\bibfnamefont{A.~N.} \bibnamefont{Kholod}},
  \bibinfo{author}{\bibfnamefont{S.}~\bibnamefont{Ossicini}},
  \bibinfo{author}{\bibfnamefont{V.~E.} \bibnamefont{Borisenko}},
  \bibnamefont{and}
  \bibinfo{author}{\bibfnamefont{F.}~\bibnamefont{Arnaud~d\char39{}Avitaya}},
  \bibinfo{journal}{Phys. Rev. B} \textbf{\bibinfo{volume}{65}},
  \bibinfo{pages}{115315} (\bibinfo{year}{2002}).

\bibitem[{\citenamefont{Kholod et~al.}(2004)\citenamefont{Kholod, Shaposhnikov,
  Sobolev, Borisenko, D\char39{}Avitaya, and Ossicini}}]{kholod-04}
\bibinfo{author}{\bibfnamefont{A.~N.} \bibnamefont{Kholod}},
  \bibinfo{author}{\bibfnamefont{V.~L.} \bibnamefont{Shaposhnikov}},
  \bibinfo{author}{\bibfnamefont{N.}~\bibnamefont{Sobolev}},
  \bibinfo{author}{\bibfnamefont{V.~E.} \bibnamefont{Borisenko}},
  \bibinfo{author}{\bibfnamefont{F.~A.} \bibnamefont{D\char39{}Avitaya}},
  \bibnamefont{and} \bibinfo{author}{\bibfnamefont{S.}~\bibnamefont{Ossicini}},
  \bibinfo{journal}{Phys. Rev. B} \textbf{\bibinfo{volume}{70}},
  \bibinfo{pages}{035317} (\bibinfo{year}{2004}).

\bibitem[{\citenamefont{Tsolakidis and Martin}(2005)}]{tsolakidis-05}
\bibinfo{author}{\bibfnamefont{A.}~\bibnamefont{Tsolakidis}} \bibnamefont{and}
  \bibinfo{author}{\bibfnamefont{R.~M.} \bibnamefont{Martin}},
  \bibinfo{journal}{Phys. Rev. B} \textbf{\bibinfo{volume}{71}},
  \bibinfo{eid}{125319} (\bibinfo{year}{2005}).

\bibitem[{\citenamefont{Melnikov and Chelikowsky}(2003)}]{melnikov-ssc}
\bibinfo{author}{\bibfnamefont{D.}~\bibnamefont{Melnikov}} \bibnamefont{and}
  \bibinfo{author}{\bibfnamefont{J.}~\bibnamefont{Chelikowsky}},
  \bibinfo{journal}{Solid State Commun.} \textbf{\bibinfo{volume}{127}},
  \bibinfo{pages}{361} (\bibinfo{year}{2003}).

\bibitem[{\citenamefont{{H.-Ch. Weissker} et~al.}(2004)\citenamefont{{H.-Ch.
  Weissker}, Furthm\"uller, and Bechstedt}}]{weissker-dscf}
\bibinfo{author}{\bibnamefont{{H.-Ch. Weissker}}},
  \bibinfo{author}{\bibfnamefont{J.}~\bibnamefont{Furthm\"uller}},
  \bibnamefont{and}
  \bibinfo{author}{\bibfnamefont{F.}~\bibnamefont{Bechstedt}},
  \bibinfo{journal}{Phys. Rev. B} \textbf{\bibinfo{volume}{69}},
  \bibinfo{eid}{115310} (\bibinfo{year}{2004}).

\bibitem[{\citenamefont{Schmidt et~al.}(2002)\citenamefont{Schmidt, Denker,
  Christiansen, and Ernst}}]{schmidt-02}
\bibinfo{author}{\bibfnamefont{O.~G.} \bibnamefont{Schmidt}},
  \bibinfo{author}{\bibfnamefont{U.}~\bibnamefont{Denker}},
  \bibinfo{author}{\bibfnamefont{S.}~\bibnamefont{Christiansen}},
  \bibnamefont{and} \bibinfo{author}{\bibfnamefont{F.}~\bibnamefont{Ernst}},
  \bibinfo{journal}{Appl. Phys. Lett.} \textbf{\bibinfo{volume}{81}},
  \bibinfo{pages}{2614} (\bibinfo{year}{2002}).

\bibitem[{\citenamefont{Takeoka et~al.}(2000)\citenamefont{Takeoka, Toshikiyo,
  Fujii, Hayashi, and Yamamoto}}]{takeoka-00-GeSi}
\bibinfo{author}{\bibfnamefont{S.}~\bibnamefont{Takeoka}},
  \bibinfo{author}{\bibfnamefont{K.}~\bibnamefont{Toshikiyo}},
  \bibinfo{author}{\bibfnamefont{M.}~\bibnamefont{Fujii}},
  \bibinfo{author}{\bibfnamefont{S.}~\bibnamefont{Hayashi}}, \bibnamefont{and}
  \bibinfo{author}{\bibfnamefont{K.}~\bibnamefont{Yamamoto}},
  \bibinfo{journal}{Phys. Rev. B} \textbf{\bibinfo{volume}{61}},
  \bibinfo{pages}{15988} (\bibinfo{year}{2000}).

\bibitem[{\citenamefont{{H.-Ch. Weissker}
  et~al.}(2003{\natexlab{c}})\citenamefont{{H.-Ch. Weissker}, Furthm\"uller,
  and Bechstedt}}]{weissker-GeSi-prl}
\bibinfo{author}{\bibnamefont{{H.-Ch. Weissker}}},
  \bibinfo{author}{\bibfnamefont{J.}~\bibnamefont{Furthm\"uller}},
  \bibnamefont{and}
  \bibinfo{author}{\bibfnamefont{F.}~\bibnamefont{Bechstedt}},
  \bibinfo{journal}{Phys. Rev. Lett.} \textbf{\bibinfo{volume}{90}},
  \bibinfo{eid}{085501} (\bibinfo{year}{2003}{\natexlab{c}}).

\bibitem[{\citenamefont{Yang et~al.}(2006)\citenamefont{Yang, Jin, Kim, and
  Jo}}]{yang-06}
\bibinfo{author}{\bibfnamefont{J.-E.} \bibnamefont{Yang}},
  \bibinfo{author}{\bibfnamefont{C.-B.} \bibnamefont{Jin}},
  \bibinfo{author}{\bibfnamefont{C.-J.} \bibnamefont{Kim}}, \bibnamefont{and}
  \bibinfo{author}{\bibfnamefont{M.-H.} \bibnamefont{Jo}},
  \bibinfo{journal}{Nano Letters} \textbf{\bibinfo{volume}{6}},
  \bibinfo{pages}{2679} (\bibinfo{year}{2006}).

\bibitem[{\citenamefont{Migas and Borisenko}(2007)}]{migas-07}
\bibinfo{author}{\bibfnamefont{D.~B.} \bibnamefont{Migas}} \bibnamefont{and}
  \bibinfo{author}{\bibfnamefont{V.~E.} \bibnamefont{Borisenko}},
  \bibinfo{journal}{Phys. Rev. B} \textbf{\bibinfo{volume}{76}},
  \bibinfo{eid}{035440} (\bibinfo{year}{2007}).

\bibitem[{\citenamefont{Delerue et~al.}(2000)\citenamefont{Delerue, Lannoo, and
  Allan}}]{delerue-00}
\bibinfo{author}{\bibfnamefont{C.}~\bibnamefont{Delerue}},
  \bibinfo{author}{\bibfnamefont{M.}~\bibnamefont{Lannoo}}, \bibnamefont{and}
  \bibinfo{author}{\bibfnamefont{G.}~\bibnamefont{Allan}},
  \bibinfo{journal}{Phys. Rev. Lett.} \textbf{\bibinfo{volume}{84}},
  \bibinfo{pages}{2457} (\bibinfo{year}{2000}).

\bibitem[{\citenamefont{Runge and Gross}(1984)}]{runge-gross-84}
\bibinfo{author}{\bibfnamefont{E.}~\bibnamefont{Runge}} \bibnamefont{and}
  \bibinfo{author}{\bibfnamefont{E.}~\bibnamefont{Gross}},
  \bibinfo{journal}{Phys. Rev. Lett.} \textbf{\bibinfo{volume}{52}},
  \bibinfo{pages}{997} (\bibinfo{year}{1984}).

\bibitem[{\citenamefont{Marques et~al.}(2006)\citenamefont{Marques, Ullrich,
  Nogueira, Rubio, Burke, and Gross}}]{pMarques2006}
\bibinfo{editor}{\bibfnamefont{M.~A.~L.} \bibnamefont{Marques}},
  \bibinfo{editor}{\bibfnamefont{C.}~\bibnamefont{Ullrich}},
  \bibinfo{editor}{\bibfnamefont{F.}~\bibnamefont{Nogueira}},
  \bibinfo{editor}{\bibfnamefont{A.}~\bibnamefont{Rubio}},
  \bibinfo{editor}{\bibfnamefont{K.}~\bibnamefont{Burke}}, \bibnamefont{and}
  \bibinfo{editor}{\bibfnamefont{E.~K.~U.} \bibnamefont{Gross}}, eds.,
  \emph{\bibinfo{title}{Time-Dependent Density Functional Theory}}, vol.
  \bibinfo{volume}{706} of \emph{\bibinfo{series}{Lecture Notes in Physics}}
  (\bibinfo{publisher}{Springer-Verlag}, \bibinfo{address}{Berlin},
  \bibinfo{year}{2006}).

\bibitem[{\citenamefont{Botti et~al.}(2007)\citenamefont{Botti, Schindlmayr,
  {Del~Sole}, and Reining}}]{botti-review-07}
\bibinfo{author}{\bibfnamefont{S.}~\bibnamefont{Botti}},
  \bibinfo{author}{\bibfnamefont{A.}~\bibnamefont{Schindlmayr}},
  \bibinfo{author}{\bibfnamefont{R.}~\bibnamefont{{Del~Sole}}},
  \bibnamefont{and} \bibinfo{author}{\bibfnamefont{L.}~\bibnamefont{Reining}},
  \bibinfo{journal}{Rep. Prog. Phys.} \textbf{\bibinfo{volume}{70}},
  \bibinfo{pages}{357} (\bibinfo{year}{2007}).

\bibitem[{\citenamefont{Marques et~al.}(2001)\citenamefont{Marques, Castro, and
  Rubio}}]{pMarques2001}
\bibinfo{author}{\bibfnamefont{M.~A.~L.} \bibnamefont{Marques}},
  \bibinfo{author}{\bibfnamefont{A.}~\bibnamefont{Castro}}, \bibnamefont{and}
  \bibinfo{author}{\bibfnamefont{A.}~\bibnamefont{Rubio}},
  \bibinfo{journal}{J.~Chem.~Phys.} \textbf{\bibinfo{volume}{115}},
  \bibinfo{pages}{3006} (\bibinfo{year}{2001}).

\bibitem[{\citenamefont{{H.-Ch. Weissker} et~al.}(2006)\citenamefont{{H.-Ch.
  Weissker}, Serrano, Huotari, Bruneval, Sottile, Monaco, Krisch, Olevano, and
  Reining}}]{weissker-ixs-prl}
\bibinfo{author}{\bibnamefont{{H.-Ch. Weissker}}},
  \bibinfo{author}{\bibfnamefont{J.}~\bibnamefont{Serrano}},
  \bibinfo{author}{\bibfnamefont{S.}~\bibnamefont{Huotari}},
  \bibinfo{author}{\bibfnamefont{F.}~\bibnamefont{Bruneval}},
  \bibinfo{author}{\bibfnamefont{F.}~\bibnamefont{Sottile}},
  \bibinfo{author}{\bibfnamefont{G.}~\bibnamefont{Monaco}},
  \bibinfo{author}{\bibfnamefont{M.}~\bibnamefont{Krisch}},
  \bibinfo{author}{\bibfnamefont{V.}~\bibnamefont{Olevano}}, \bibnamefont{and}
  \bibinfo{author}{\bibfnamefont{L.}~\bibnamefont{Reining}},
  \bibinfo{journal}{Phys. Rev. Lett.} \textbf{\bibinfo{volume}{97}},
  \bibinfo{eid}{237602} (\bibinfo{year}{2006}).

\bibitem[{\citenamefont{Onida et~al.}(2002)\citenamefont{Onida, Reining, and
  Rubio}}]{rmp_lucia}
\bibinfo{author}{\bibfnamefont{G.}~\bibnamefont{Onida}},
  \bibinfo{author}{\bibfnamefont{L.}~\bibnamefont{Reining}}, \bibnamefont{and}
  \bibinfo{author}{\bibfnamefont{A.}~\bibnamefont{Rubio}},
  \bibinfo{journal}{Rev. Mod. Phys.} \textbf{\bibinfo{volume}{74}},
  \bibinfo{pages}{601} (\bibinfo{year}{2002}), \bibinfo{note}{and references
  therein}.

\bibitem[{\citenamefont{{H.-Ch. Weissker}
  et~al.}(2003{\natexlab{d}})\citenamefont{{H.-Ch. Weissker}, Furthm\"uller,
  and Bechstedt}}]{weissker-emt}
\bibinfo{author}{\bibnamefont{{H.-Ch. Weissker}}},
  \bibinfo{author}{\bibfnamefont{J.}~\bibnamefont{Furthm\"uller}},
  \bibnamefont{and}
  \bibinfo{author}{\bibfnamefont{F.}~\bibnamefont{Bechstedt}},
  \bibinfo{journal}{Phys. Rev. B} \textbf{\bibinfo{volume}{67}},
  \bibinfo{eid}{165322} (\bibinfo{year}{2003}{\natexlab{d}}).

\bibitem[{web()}]{web-vasp}
\bibinfo{note}{{h}ttp://cms.mpi.univie.ac.at/vasp/}.

\bibitem[{\citenamefont{Perdew and Zunger}(1981)}]{perdew-zunger}
\bibinfo{author}{\bibfnamefont{J.~P.} \bibnamefont{Perdew}} \bibnamefont{and}
  \bibinfo{author}{\bibfnamefont{A.}~\bibnamefont{Zunger}},
  \bibinfo{journal}{Phys. Rev. B} \textbf{\bibinfo{volume}{23}},
  \bibinfo{pages}{5048} (\bibinfo{year}{1981}).

\bibitem[{\citenamefont{Bl\"ochl}(1994)}]{bloechl-94}
\bibinfo{author}{\bibfnamefont{P.~E.} \bibnamefont{Bl\"ochl}},
  \bibinfo{journal}{Phys. Rev. B} \textbf{\bibinfo{volume}{50}},
  \bibinfo{pages}{17953} (\bibinfo{year}{1994}).

\bibitem[{\citenamefont{Marques et~al.}(2003)\citenamefont{Marques, Castro,
  Bertsch, and Rubio}}]{octopus}
\bibinfo{author}{\bibfnamefont{M.~A.~L.} \bibnamefont{Marques}},
  \bibinfo{author}{\bibfnamefont{A.}~\bibnamefont{Castro}},
  \bibinfo{author}{\bibfnamefont{G.~F.} \bibnamefont{Bertsch}},
  \bibnamefont{and} \bibinfo{author}{\bibfnamefont{A.}~\bibnamefont{Rubio}},
  \bibinfo{journal}{Comp. Phys. Comm.} \textbf{\bibinfo{volume}{151}},
  \bibinfo{pages}{60} (\bibinfo{year}{2003}).

\bibitem[{\citenamefont{Castro et~al.}(2006)\citenamefont{Castro, Marques,
  Appel, Oliveira, Rozzi, Andrade, Lorenzen, Gross, and Rubio}}]{octopus2}
\bibinfo{author}{\bibfnamefont{A.}~\bibnamefont{Castro}},
  \bibinfo{author}{\bibfnamefont{M.~A.~L.} \bibnamefont{Marques}},
  \bibinfo{author}{\bibfnamefont{H.}~\bibnamefont{Appel}},
  \bibinfo{author}{\bibfnamefont{M.}~\bibnamefont{Oliveira}},
  \bibinfo{author}{\bibfnamefont{C.}~\bibnamefont{Rozzi}},
  \bibinfo{author}{\bibfnamefont{X.}~\bibnamefont{Andrade}},
  \bibinfo{author}{\bibfnamefont{F.}~\bibnamefont{Lorenzen}},
  \bibinfo{author}{\bibfnamefont{E.~K.~U.} \bibnamefont{Gross}},
  \bibnamefont{and} \bibinfo{author}{\bibfnamefont{A.}~\bibnamefont{Rubio}},
  \bibinfo{journal}{Phys. Stat. Sol. (b)} \textbf{\bibinfo{volume}{243}},
  \bibinfo{pages}{2465} (\bibinfo{year}{2006}).

\bibitem[{\citenamefont{Troullier and Martins}(1991)}]{troullier91}
\bibinfo{author}{\bibfnamefont{N.}~\bibnamefont{Troullier}} \bibnamefont{and}
  \bibinfo{author}{\bibfnamefont{J.}~\bibnamefont{Martins}},
  \bibinfo{journal}{Phys. Rev. B} \textbf{\bibinfo{volume}{43}},
  \bibinfo{pages}{1993} (\bibinfo{year}{1991}).

\bibitem[{\citenamefont{Castro et~al.}(2004)\citenamefont{Castro, Marques,
  Alonso, and Rubio}}]{castro04}
\bibinfo{author}{\bibfnamefont{A.}~\bibnamefont{Castro}},
  \bibinfo{author}{\bibfnamefont{M.~A.~L.} \bibnamefont{Marques}},
  \bibinfo{author}{\bibfnamefont{J.~A.} \bibnamefont{Alonso}},
  \bibnamefont{and} \bibinfo{author}{\bibfnamefont{A.}~\bibnamefont{Rubio}},
  \bibinfo{journal}{J. Comp. Theoret. Nanoscience}
  \textbf{\bibinfo{volume}{1}}, \bibinfo{pages}{231} (\bibinfo{year}{2004}).

\bibitem[{\citenamefont{Casida}(1995)}]{casida-95}
\bibinfo{author}{\bibfnamefont{M.~E.} \bibnamefont{Casida}}, in
  \emph{\bibinfo{booktitle}{Recent Advances in Density Functional Methods, {\rm
  Part I}}}, edited by \bibinfo{editor}{\bibfnamefont{D.}~\bibnamefont{Chong}}
  (\bibinfo{publisher}{World Scientific}, \bibinfo{year}{1995}), p.
  \bibinfo{pages}{155}.

\bibitem[{\citenamefont{Casida}(1996)}]{casida-96}
\bibinfo{author}{\bibfnamefont{M.~E.} \bibnamefont{Casida}}, in
  \emph{\bibinfo{booktitle}{Recent Developments and Applications of Modern
  Density Functional Theory}}, edited by
  \bibinfo{editor}{\bibfnamefont{J.}~\bibnamefont{Seminario}}
  (\bibinfo{publisher}{Elsevier Science, Amsterdam}, \bibinfo{year}{1996}), p.
  \bibinfo{pages}{391}.

\bibitem[{\citenamefont{Adolph et~al.}(2001)\citenamefont{Adolph,
  Furthm\"uller, and Bechstedt}}]{adolph-01}
\bibinfo{author}{\bibfnamefont{B.}~\bibnamefont{Adolph}},
  \bibinfo{author}{\bibfnamefont{J.}~\bibnamefont{Furthm\"uller}},
  \bibnamefont{and}
  \bibinfo{author}{\bibfnamefont{F.}~\bibnamefont{Bechstedt}},
  \bibinfo{journal}{Phys. Rev. B} \textbf{\bibinfo{volume}{63}},
  \bibinfo{pages}{125108} (\bibinfo{year}{2001}).

\bibitem[{\citenamefont{{H.-Ch. Weissker} et~al.}(2001)\citenamefont{{H.-Ch.
  Weissker}, Furthm\"uller, and Bechstedt}}]{weissker-tetra}
\bibinfo{author}{\bibnamefont{{H.-Ch. Weissker}}},
  \bibinfo{author}{\bibfnamefont{J.}~\bibnamefont{Furthm\"uller}},
  \bibnamefont{and}
  \bibinfo{author}{\bibfnamefont{F.}~\bibnamefont{Bechstedt}},
  \bibinfo{journal}{Phys. Rev. B} \textbf{\bibinfo{volume}{64}},
  \bibinfo{eid}{035105} (\bibinfo{year}{2001}).

\bibitem[{\citenamefont{Degoli et~al.}(2004)\citenamefont{Degoli, Cantele,
  Luppi, Magri, Ninno, Bisi, and Ossicini}}]{degoli-04}
\bibinfo{author}{\bibfnamefont{E.}~\bibnamefont{Degoli}},
  \bibinfo{author}{\bibfnamefont{G.}~\bibnamefont{Cantele}},
  \bibinfo{author}{\bibfnamefont{E.}~\bibnamefont{Luppi}},
  \bibinfo{author}{\bibfnamefont{R.}~\bibnamefont{Magri}},
  \bibinfo{author}{\bibfnamefont{D.}~\bibnamefont{Ninno}},
  \bibinfo{author}{\bibfnamefont{O.}~\bibnamefont{Bisi}}, \bibnamefont{and}
  \bibinfo{author}{\bibfnamefont{S.}~\bibnamefont{Ossicini}},
  \bibinfo{journal}{Phys. Rev. B} \textbf{\bibinfo{volume}{69}},
  \bibinfo{eid}{155411} (\bibinfo{year}{2004}).

\bibitem[{\citenamefont{Maxwell-Garnett}(1904)}]{emt}
\bibinfo{author}{\bibfnamefont{J.}~\bibnamefont{Maxwell-Garnett}},
  \bibinfo{journal}{Philos. Trans. R. Soc.} \textbf{\bibinfo{volume}{203}},
  \bibinfo{pages}{305} (\bibinfo{year}{1904}).

\bibitem[{\citenamefont{Wood and Ashcroft}(1977)}]{emt-b}
\bibinfo{author}{\bibfnamefont{D.~M.} \bibnamefont{Wood}} \bibnamefont{and}
  \bibinfo{author}{\bibfnamefont{N.~W.} \bibnamefont{Ashcroft}},
  \bibinfo{journal}{Philos. Mag.} \textbf{\bibinfo{volume}{35}},
  \bibinfo{pages}{269} (\bibinfo{year}{1977}).

\bibitem[{\citenamefont{Sottile et~al.}(2005)\citenamefont{Sottile, Bruneval,
  Marinopoulos, Dash, Botti, Olevano, Vast, Rubio, and Reining}}]{sottile05}
\bibinfo{author}{\bibfnamefont{F.}~\bibnamefont{Sottile}},
  \bibinfo{author}{\bibfnamefont{F.}~\bibnamefont{Bruneval}},
  \bibinfo{author}{\bibfnamefont{A.~G.} \bibnamefont{Marinopoulos}},
  \bibinfo{author}{\bibfnamefont{L.~K.} \bibnamefont{Dash}},
  \bibinfo{author}{\bibfnamefont{S.}~\bibnamefont{Botti}},
  \bibinfo{author}{\bibfnamefont{V.}~\bibnamefont{Olevano}},
  \bibinfo{author}{\bibfnamefont{N.}~\bibnamefont{Vast}},
  \bibinfo{author}{\bibfnamefont{A.}~\bibnamefont{Rubio}}, \bibnamefont{and}
  \bibinfo{author}{\bibfnamefont{L.}~\bibnamefont{Reining}},
  \bibinfo{journal}{Int. J. Quantum Chem.} \textbf{\bibinfo{volume}{102}},
  \bibinfo{pages}{684} (\bibinfo{year}{2005}).

\bibitem[{\citenamefont{Botti et~al.}(2002)\citenamefont{Botti, Olevano,
  Reining, and Andreani}}]{botti02}
\bibinfo{author}{\bibfnamefont{S.}~\bibnamefont{Botti}},
  \bibinfo{author}{\bibfnamefont{N.~V.~V.} \bibnamefont{Olevano}},
  \bibinfo{author}{\bibfnamefont{L.}~\bibnamefont{Reining}}, \bibnamefont{and}
  \bibinfo{author}{\bibfnamefont{L.~C.} \bibnamefont{Andreani}},
  \bibinfo{journal}{Phys. Rev. Lett.} \textbf{\bibinfo{volume}{89}},
  \bibinfo{pages}{216803} (\bibinfo{year}{2002}).

\bibitem[{\citenamefont{Botti et~al.}(2004)\citenamefont{Botti, Olevano,
  Reining, and Andreani}}]{botti04}
\bibinfo{author}{\bibfnamefont{S.}~\bibnamefont{Botti}},
  \bibinfo{author}{\bibfnamefont{N.~V.~V.} \bibnamefont{Olevano}},
  \bibinfo{author}{\bibfnamefont{L.}~\bibnamefont{Reining}}, \bibnamefont{and}
  \bibinfo{author}{\bibfnamefont{L.~C.} \bibnamefont{Andreani}},
  \bibinfo{journal}{Phys. Rev. B} \textbf{\bibinfo{volume}{70}},
  \bibinfo{pages}{045301} (\bibinfo{year}{2004}).

\bibitem[{\citenamefont{{Huml\'{\i}\v{c}ek}
  et~al.}(1989)\citenamefont{{Huml\'{\i}\v{c}ek}, Garriga, Alonso, and
  Cardona}}]{humlicek-89}
\bibinfo{author}{\bibfnamefont{J.}~\bibnamefont{{Huml\'{\i}\v{c}ek}}},
  \bibinfo{author}{\bibfnamefont{M.}~\bibnamefont{Garriga}},
  \bibinfo{author}{\bibfnamefont{M.~I.} \bibnamefont{Alonso}},
  \bibnamefont{and} \bibinfo{author}{\bibfnamefont{M.}~\bibnamefont{Cardona}},
  \bibinfo{journal}{J. Appl. Phys.} \textbf{\bibinfo{volume}{65}},
  \bibinfo{pages}{2827} (\bibinfo{year}{1989}).

\bibitem[{\citenamefont{{H.-Ch. Weissker} et~al.}(2009)\citenamefont{{H.-Ch.
  Weissker}, Olevano, and Reining}}]{weissker-interpolation}
\bibinfo{author}{\bibnamefont{{H.-Ch. Weissker}}},
  \bibinfo{author}{\bibfnamefont{V.}~\bibnamefont{Olevano}}, \bibnamefont{and}
  \bibinfo{author}{\bibfnamefont{L.}~\bibnamefont{Reining}},
  \bibinfo{journal}{Phys. Rev. B, in press}  (\bibinfo{year}{2009}).

\bibitem[{\citenamefont{Tognini et~al.}(1996)\citenamefont{Tognini, Andreani,
  Geddo, Stella, Cheyssac, Kofman, and Migliori}}]{tognini-96}
\bibinfo{author}{\bibfnamefont{P.}~\bibnamefont{Tognini}},
  \bibinfo{author}{\bibfnamefont{L.~C.} \bibnamefont{Andreani}},
  \bibinfo{author}{\bibfnamefont{M.}~\bibnamefont{Geddo}},
  \bibinfo{author}{\bibfnamefont{A.}~\bibnamefont{Stella}},
  \bibinfo{author}{\bibfnamefont{P.}~\bibnamefont{Cheyssac}},
  \bibinfo{author}{\bibfnamefont{R.}~\bibnamefont{Kofman}}, \bibnamefont{and}
  \bibinfo{author}{\bibfnamefont{A.}~\bibnamefont{Migliori}},
  \bibinfo{journal}{Phys. Rev. B} \textbf{\bibinfo{volume}{53}},
  \bibinfo{pages}{6992} (\bibinfo{year}{1996}).

\bibitem[{\citenamefont{Dexter}(1958)}]{dexter-58}
\bibinfo{author}{\bibfnamefont{D.}~\bibnamefont{Dexter}},
  \bibinfo{journal}{Solid State Physics -- Advances in Research and
  Applications} \textbf{\bibinfo{volume}{6}}, \bibinfo{pages}{353}
  (\bibinfo{year}{1958}).

\bibitem[{\citenamefont{Delerue et~al.}(1993)\citenamefont{Delerue, Allan, and
  Lannoo}}]{delerue-93}
\bibinfo{author}{\bibfnamefont{C.}~\bibnamefont{Delerue}},
  \bibinfo{author}{\bibfnamefont{G.}~\bibnamefont{Allan}}, \bibnamefont{and}
  \bibinfo{author}{\bibfnamefont{M.}~\bibnamefont{Lannoo}},
  \bibinfo{journal}{Phys. Rev. B} \textbf{\bibinfo{volume}{48}},
  \bibinfo{pages}{11024} (\bibinfo{year}{1993}).

\bibitem[{\citenamefont{van Roosbroeck and Shockley}(1954)}]{vRS}
\bibinfo{author}{\bibfnamefont{W.}~\bibnamefont{van Roosbroeck}}
  \bibnamefont{and} \bibinfo{author}{\bibfnamefont{W.}~\bibnamefont{Shockley}},
  \bibinfo{journal}{Phys. Rev.} \textbf{\bibinfo{volume}{94}},
  \bibinfo{pages}{1558} (\bibinfo{year}{1954}).

\end{thebibliography}

\end{document}